\documentclass[aps,prl,superscriptaddress, twocolumn]{revtex4-1}
\usepackage[english]{babel}
\usepackage[ansinew]{inputenc}
\usepackage{times}
\usepackage{graphicx}
\usepackage{graphics}
\usepackage{amsmath}
\usepackage{amsfonts}
\usepackage{amssymb}
\usepackage{dsfont}
\usepackage{epstopdf}
\usepackage{makeidx}
\usepackage{subfigure}
\usepackage{color}
\usepackage{pgf}
\usepackage{bm}
\usepackage{ulem}
\usepackage{tikz} 

\newcommand{\be}{\begin{equation}}
\newcommand{\ee}{\end{equation}}
\newcommand{\bea}{\begin{eqnarray}}
\newcommand{\eea}{\end{eqnarray}}

\newcommand\redsout{\bgroup\markoverwith{\textcolor{red}{\rule[0.5ex]{4pt}{0.8pt}}}\ULon}

\makeindex

\begin{document}

\title{Controlling spontaneous emission via electronic correlations in transparent metals}

\author{M. B. Silva Neto} 

\affiliation{Instituto de Fisica, Universidade Federal do Rio de Janeiro, Caixa Postal
68528, Rio de Janeiro, Brazil}

\author{F. A. Pinheiro} 

\affiliation{Instituto de Fisica, Universidade Federal do Rio de Janeiro, Caixa Postal
68528, Rio de Janeiro, Brazil}

\begin{abstract}

We study the spontaneous emission of agglomerates of two-level quantum emitters embedded in 
a correlated transparent metal. The characteristic emission energy corresponds to the splitting 
between ground and excited states of a neutral, nonmagnetic molecular impurity (F color 
center), while correlations are due to the existence of narrow bands in the metal. This is 
the case of transition metal oxides with an ABO$_3$ Perovskite structure, such as SrVO$_{3}$ 
and CaVO$_{3}$, where oxygen vacancies are responsible for the emission of visible light, while 
the correlated metallic nature arises from the partial filling of a band with mostly $d$-orbital 
character. For these systems we put forward an interdisciplinary, tunable mechanism to control  
light emission governed by electronic correlations. We show that not only there exists a critical 
value for the correlation strength above which the metal becomes transparent in the visible, 
but also that strong correlations can lead to a remarkable enhancement of the light-matter 
coupling. By unveiling the role of electronic correlations in spontaneous emission, our findings set the 
basis for the design of controllable, solid-state, single-photon sources in correlated transparent metals.  

\end{abstract}

\maketitle

Progresses in the understanding of light-matter interactions have enabled new 
mechanisms of controlling light emission, propagation and extraction on chip~\cite{lipson,lodahl}. 
Since the pioneering work of Purcell it has been known that spontaneous emission 
is not only determined by the emitters' (atoms, molecules, and quantum dots) intrinsic 
electronic levels but is also influenced by the surrounding electromagnetic environment
~\cite{Purcell-46}. This discovery has paved the way to the development of several 
different strategies to modify and tailor the spontaneous emission of quantum emitters, 
typically relying on different choices of material platforms and/or physical mechanisms that affect 
spontaneous emission. Photonic crystals~\cite{goban}, dielectric microcavities~\cite{senellart}, 
nanophotonic waveguides~\cite{arcari}, graphene-based structures~\cite{tielrooij,wilton2015}, 
dielectric~\cite{regmi} and plasmonic~\cite{tanaka,muskens} nanoantennas, and 
metamaterials~\cite{noginov,roth,liu} are examples of different material platforms that have 
been explored to that purpose. In particular, metamaterials and 2D metasurfaces are 
versatile systems to achieving far-field emission patterns with desired properties, such as 
divergence and directionality~\cite{bohn}, and even completely suppressing light 
emission~\cite{wilton2013}. Metallic structures have also been employed to modify 
spontaneous emission, with plasmonic resonances being designed to increase the 
electric field and enhance the local density of states at the emitters location~\cite{pelton,makarova}. 
Regarding the different physical mechanisms to modify light emission one may cite the 
Fano effect~\cite{livro-tiago,tiago2017} and critical phenomena. Indeed there are evidences, 
both theoretical~\cite{juanjo1,juanjo2,szilard} and experimental~\cite{carminati}, that 
phase transitions affect spontaneous emission in a crucial way, thus allowing for the optical 
determination of critical exponents via the Purcell factor~\cite{mbsn}. 
Altogether these different routes have allowed for many technological applications, such as solar 
cells~\cite{oregan}, molecular imaging~\cite{moerner,vallee}, and single-photon sources~\cite{michler}. 

In the present letter we propose a twofold, interdisciplinary strategy to control spontaneous 
emission that combines: i) an alternative material platform, transition metal oxides; 
and ii) a novel physical mechanism, electronic correlations. More specifically, we consider color 
center agglomerates incorporated into a correlated transparent metal, namely ABO$_3$ Perovskites, 
such as SrVO$_{3}$ and CaVO$_{3}$, where oxygen vacancies are responsible for
emission of visible light~\cite{zhang2016} and electronic correlations arise from the 
narrowness of the partially filled B- ion, $d$-bands. Our choice for this particular 
material platform is motivated by the fact that transition metal oxides exhibit unique electrical and 
optical properties, such as excellent carrier mobilities, mechanical stress tolerance, compatibility 
with organic dielectric and photoactive materials, and high optical transparency~\cite{xu}. These 
materials are also versatile and cost-effective to many applications in optoelectronics, such as electronic 
circuits, flexible organic light-emitting diode (OLED) displays, and solar cells~\cite{xu}. 

%
\begin{figure}
\includegraphics[scale=0.4]{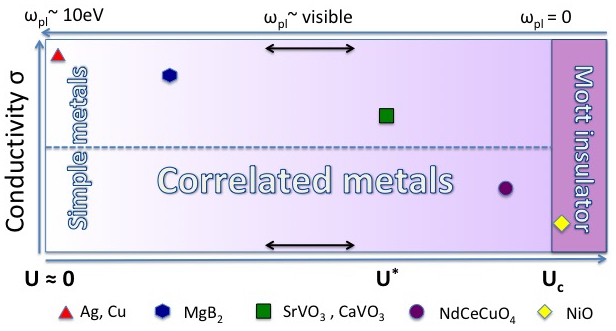}
\caption{Schematic evolution of the conductivity, $\sigma$, and plasmon
frequency, $\omega_{pl}$, as a function of the correlation strength, $U$. 
While simple metals like Ag and Cu are excellent conductors (large 
$\sigma$) and highly reflective (large $\omega_{pl}$), correlated metals 
such as SrVO$_3$ and CaVO$_3$ exhibit conductivities that are still higher 
than the best doped semiconductors  (dashed line) while transmitting 
visible light very effectively as they have $\omega_{pl}$ below the 
visible transparency window.}
\label{Fig-Correlated-Metals}
\end{figure}
%

A key issue while engineering metallic materials for optoelectronic applications 
relies on an adequate combination of high optical transparency and high electrical conductivity. 
This is usually challenging since the plasmon frequency of the best conducting metals is 
typically of the order of $\hbar\omega_{pl}\sim10eV$, well above the visible transparency window 
$1.8eV\leq\hbar\omega\leq3.1eV$, see Fig. \ref{Fig-Correlated-Metals}. Hence achieving 
transparency in good conductors requires minimizing the plasmon frequency $\omega_{pl}^{2}=e^{2}n/\varepsilon_0 m^*$ 
(given in terms of the electric charge $e$, the dielectric constant in free space $\varepsilon_{0}$, 
the electronic density in the metal $n$, of the effective mass $m^{*}$), while keeping the electrical 
conductivity $\sigma = e^{2} \tau (n/m^{*})$ (with $\tau$ the scattering time) large enough by a 
judicious choice of the ratio $n/m^{*}$. The tradicional strategy to adjust the ratio $n/m^{*}$ relies 
on a trade-off between increasing $n$, as in wide-bandgap 
semiconductors, and/or decreasing $m^*$ via heavy doping, as in 
the case of tin-doped indium oxide (ITO). Nevertheless, although ITO exhibits 
the largest conductivity for materials of its class (horizontal dashed line in Fig. \ref{Fig-Correlated-Metals}), it is 
still much smaller than the conductivities of transition
metals oxides, such as SrVO$_3$ and CaVO$_3$. 
These materials have very good transmission efficiency of the order of 
$80\%$, except in the blue region \cite{zhang2016}, besides being excellent conductors (see 
Fig. \ref{Fig-Correlated-Metals}), remarkable properties on which we capitalize to achieve 
enhanced and controllable spontaneous emission.

In what follows we propose for the first time an alternative physical mechanism, based on 
electronic correlations, for: i) reducing the plasmon frequency of transition 
metal oxides, making them transparent in the visible; and ii) controlling and enhancing the 
spontaneous emission of visible light. We start by recalling that moderate to strong electronic 
correlations, represented by the Hubbard interaction $U$, renormalizes the effective mass 
of carriers according to $m^{*}(U)/m^*(0)=1/[1-\left(U/U_{c}\right)^{2}]$, 
where $U_c\approx 5$eV is the critical value for the metal-to-Mott-insulator transition
in these systems \cite{Brinkmann-Rice}. The plasmon frequency then modifies as
\begin{equation}
\omega_{pl}^{2}\left(U\right)=\omega_{pl}^{2}\left(0\right)\left[1-\left(U/U_{c}\right)^{2}\right].
\end{equation}
%
Such renormalization has indeed been observed in SrVO$_3$ by ARPES \cite{SrVO3-ARPES}, 
where $U\approx 3.2$eV, and $\hbar\omega_{pl}(U)< 1.33$eV, 
{\it i.e.} below the lower visible edge \cite{zhang2016}, see Fig. \ref{Fig-Correlated-Metals}.

%
\begin{figure}
\includegraphics[scale=0.35]{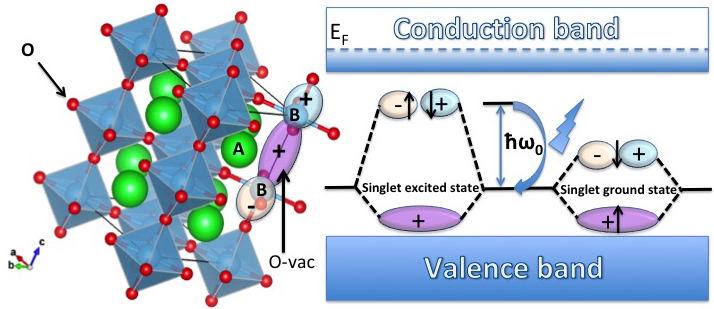}
\caption{Left: crystal structure of ABO$_3$ perovskites with an oxygen vacancy. 
Each vacancy traps two electrons in a singlet that fill up the bonding and 
anti-bonding molecular orbitals that result from the hybridization between the $e_g$
wavefunctions of the two neighbouring B-ions. Right: schematic band structure 
of ABO$_3$ perovskites with O vacancy impurity states. Spontaneous emission of in-gap color 
centers in such correlated metals occurs when the singlet excited state relaxes to 
the singlet ground state by emitting light in the visible.}
\label{Fig-Crystal-Structure-O-Vac}
\end{figure}
%

Having established the transparency of our material platform, in the following we specify
the properties of the two-level quantum emitter. The crystal structure of ABO$_3$ 
perovskites consists of corner shared BO$_6$ octahedra, with the transition metal 
B, inside each octahedron, and with the cation A, at the center of a unit cell of coordination 12 ( 
Fig. \ref{Fig-Crystal-Structure-O-Vac}). The removal of one oxygen atom from the 
structure, {\it i.e.} the introduction of an oxygen vacancy (O-vac), causes the trapping 
of two electrons, at the vacancy, each coming from a neaby B-ion. This scenario is consistent 
with LDA+DMFT calculations on oxygen deficient supercells in SrVO$_3$, and has 
been confirmed by ARPES for the case of UV irradiated SrVO$_3$ crystals 
\cite{SrVO3-ARPES-II}. The large Coulomb repulsion imposes that the spins of these electrons be anti-parallel due to virtual 
tunneling processes. As a result, the ground state of an O-vac
is a neutral, nonmagnetic spin-single state (an F color center), where the two 
electrons fill up the two molecular orbitals (bonding and anti-bonding) obtained 
from the hybridization of the original $e_g$ orbitals of the neighboring 
B ions, Fig. \ref{Fig-Crystal-Structure-O-Vac}. 

The electronic structure at the O-vac discussed above can be calculated by the 
following two-site Hamiltonian
\begin{eqnarray}
H&=&\sum_{i=1,2}\sum_{\sigma=\uparrow,\downarrow}{\cal E}_0 n_{i,\sigma}-
t\sum_{\sigma=\uparrow,\downarrow}(c^\dagger_{1,\sigma}c_{2,\sigma}+h.c.)\nonumber\\
&+&U\sum_{i=1,2}\sum_{\sigma=\uparrow,\downarrow}n_{i,\sigma}n_{i,-\sigma}+
V\sum_{\sigma,\sigma^\prime}n_{1,\sigma}n_{2,\sigma^\prime}\nonumber,
\end{eqnarray}
where ${\cal E}_0$ is the onsite energy, $t$ is the direct tunnelling, $U$ is the 
onsite Coulomb repulsion that opposes double occupancy, $V$ is the 
nearest neighbour Coulomb repulsion that opposes direct tunneling, 
and $n_{i,\sigma}=c^\dagger_{i,\sigma}c_{i,\sigma}$ 
with $c^\dagger_{i,\sigma}$ and $c_{i,\sigma}$ corresponding to the creation 
and anihilation operators for the electrons at the two B-ion, $d$-orbitals. A mean 
field treatment of the Coulomb interactions allows us to replace 
$\sum_{\sigma=\uparrow,\downarrow}\left[{\cal E}_0 n_{i,\sigma}+Un_{i,\sigma}n_{i,-\sigma}\right]$
by $\sum_{\sigma=\uparrow,\downarrow}{\cal E}_{i,\sigma} n_{i,\sigma}$, in terms of
${\cal E}_{i,\sigma}={\cal E}_0+U\langle n_{i,-\sigma}\rangle$, and also to replace
$\sum_{\sigma,\sigma^\prime}[t \delta_{\sigma,\sigma^\prime} c^\dagger_{1,\sigma}c_{2,\sigma^\prime}-V n_{1,\sigma}n_{2,\sigma^\prime}]$
by $\sum_{\sigma=\uparrow,\downarrow} {\cal T}_{\sigma} c^\dagger_{1,\sigma}c_{2,\sigma}$,
in terms of ${\cal T}_{\sigma}=t-V\sum_{\sigma^\prime}\langle c_{1,\sigma} c^\dagger_{2,\sigma^\prime}\rangle$. 
The Hamiltonian is now quadratic and can be diagonalized providing eigenvalues
\begin{equation}
E^{\pm}_\sigma=\frac{{\cal E}_{1,\sigma}+{\cal E}_{2,\sigma}\pm\sqrt{({\cal E}_{1,\sigma}-{\cal E}_{2,\sigma})^2+4{\cal T}_\sigma^2}}{2}.
\label{Eigenvalues}
\end{equation}
The Hilbert space for the O-vac problem is spanned by four basis states
$\left|1,\uparrow\downarrow;2,0\right>, \left|1,\uparrow;2,\downarrow\right>,\left|1,\downarrow;2,\uparrow\right>,\left|1,0;2,\uparrow\downarrow\right>$,
in terms of which 
$\langle n_{i,-\sigma}\rangle$ and $\langle c_{1,\sigma} c^\dagger_{2,\sigma^\prime}\rangle$
must be calculated. If the two electrons occupy the same molecular orbital,
$\langle n_{i,-\sigma}\rangle=1/2$ (projected subspace of doubly-occupied states), 
whereas if the two electrons occupy a different molecular orbital each, 
$\langle n_{i,-\sigma}\rangle=0$ (projected subspace of singly-occupied states).
Analogously,
$\langle c_{1,\sigma} c^\dagger_{2,\sigma^\prime}\rangle=\delta_{\sigma,\sigma^\prime}/2$
(projected subspace of mixed singly- and double-occupied states).
For $U>2(2t-V)$ the ground state $|g\rangle$, corresponds to a singlet configuration 
with the two electrons occupying a different molecular orbital each. There is a low lying 
singlet exited state $|le\rangle$, corresponding to a doubly-occupied bonding orbital, and 
a high energy singlet excited state $|he\rangle$, corresponding to a doubly-occupied 
anti-bonding orbital. The $|g\rangle$ and $|he\rangle$ states are shown in 
Fig. \ref{Fig-Crystal-Structure-O-Vac} and their splitting is
\begin{equation}
\hbar\omega_0 = 2t-V + \frac{U}{2}.
\end{equation}
By using typical values for the hopping, $t\approx 1.0$eV, and Hubbard parameters, 
$U\approx 3.2$eV, and $V\approx 0.8$eV, consistent with GW+DMFT and LDA+DMFT 
calculations for SrVO$_3$ \cite{LDA-DMFT-SrVO3}, we obtain 
$\hbar\omega_0\approx 2.8$eV for the characteristic emitter's energy, which 
corresponds to light emission/absorption in the blue, exactly where SrVO$_3$ 
transmits poorly \cite{zhang2016}. Similar values for $\hbar\omega_0$ 
can also be found for other ABO$_3$ Perovskites, using similar values
for $t,U,V$, satisfying $V<t<U$, all within the visible,
as observed in photoluminescence experiments in a variety of disordered 
Perovskites such as BaTiO$_3$, CaTiO$_3$, PbTiO$_3$, LiNbO$_3$, SrWO$_4$,
besides SrVO$_3$ itself \cite{PL-Disordered-ABO3}. 

%
\begin{figure}
\includegraphics[scale=0.35]{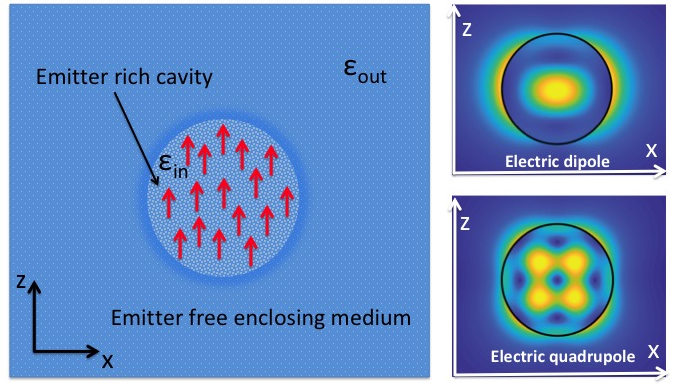}
\caption{Left: Oxygen vacancy rich, spherical cavity of dielectric constant 
$\varepsilon_{in}$, containing an agglomerate of electric dipoles (red
arrows), inside an otherwise O-vac free enclosing host of dielectric constant 
$\varepsilon_{out}$. Right: normal electromagnetic modes inside the cavity according 
to Mie's theory; only the electric dipole (top) and quadrupole (bottom) 
modes are shown for clarity.}
\label{Fig-Local-Fields-Cavity}
\end{figure}
%

We are now ready to calculate the spontaneous emission rate $\Gamma$ for a dilute and
homogeneous collection of O-vac impurity states embedded in a correlated metal 
whose optical properties depend on the correlation 
strength $U$ through the dielectric function $\varepsilon(\omega,U)=\varepsilon_0\epsilon_r(\omega,U)$. 
To this end we calculate the electric-dipole matrix element between the singlet, ground $|g\rangle$, and 
highest excited $|he\rangle$ states
\be
\left|\langle g\left|H_{int}\right|he\rangle\right|=
\sqrt{\frac{\hbar\omega_0}{2\varepsilon_{0}\epsilon_{r}\left(\omega_0,U\right) V}}\hat{\epsilon}_{\lambda}\cdot\langle g|\vec{\mu}|he\rangle,
\ee
describing the coupling between an O-vacancy, with nonzero electric-dipole moment $\vec{\mu}$ 
between the bonding and anti-bonding orbitals, and radiation with polarization 
$\hat{\epsilon_{\lambda}}$. Here $V$ is the volume and
$\epsilon_{r}\left(\omega,U\right)=\epsilon_{1}\left(\omega,U\right)+i\epsilon_{2}\left(\omega,U\right)$, with
\bea
\epsilon_{1}\left(\omega,U\right)&=&1-\frac{\omega_{pl}^2(U)}{\omega^2+\gamma^2},\nonumber\\
\epsilon_{2}\left(\omega,U\right)&=&\frac{\gamma\omega_{pl}^2(U)}{\omega(\omega^2+\gamma^2)},
\eea
is the relative permitivity of a lossy, $\gamma\neq 0$, medium. For isotropic systems with electric dipole 
moment $\vec{\mu}=\mu\hat{z}$, we can write 
$\left|\hat{\epsilon}_{\lambda}\cdot\langle g|\vec{\mu}|he\rangle\right|^{2}=
\mu^{2}\left|\left\langle \hat{\epsilon}_{\lambda}\cdot\hat{z}\right\rangle \right|^{2}=
\mu^{2}/3$. By defining the quantity $g^{2}=\hbar\omega_{0}\mu^{2}/6\varepsilon_{0}$, 
and recalling the definition of 
the spectral distribution of electromagnetic modes in the medium, 
${\cal{A}}_U(\omega,|{\bf k}|)=-(2c|{\bf k}|/\pi){\cal{I}}m[G^{R}_U\left(\omega,{\bf k}\right)]$, where 
retarded photon propagator is \cite{AGD}
\begin{equation}
G^R_{U}\left(\omega,{\bf k}\right)=\frac{1}{\epsilon_{1}\left(\omega,U\right)\omega^{2}-c^{2}|{\bf k}|^2+i\epsilon_{2}\left(\omega,U\right)\omega^{2}},
\end{equation}
we can use Fermi's golden rule to calculate the spontaneous emission rate for the $i$-th isolated,
single O-vac, in terms of the value in free space, $\Gamma_0=(2\pi/\hbar^2)g^2 [\omega_0^2/\pi^2 c^3]$, as
\begin{widetext}
\be
\Gamma_{i}(U)=\left(\frac{2\pi}{\hbar^2}\right)\left[\frac{g^{2}}{\epsilon_{r}\left(\omega_{0},U\right)}\right]\int_{0}^{\infty}\frac{dk}{\pi^2}\:k^{2}
\left\{\frac{2ck\,\epsilon_{2}\left(\omega,U\right)\omega^{2}}{\left[\epsilon_{1}\left(\omega,U\right)\omega^{2}-c^{2}k^{2}\right]^{2}+
\epsilon_{2}^{2}\left(\omega,U\right)\omega^{4}}\right\}=\eta(\omega_0,U)\Gamma_0.
\ee
\end{widetext}
This result generalizes the one obtained for the decay rate of excited atoms in absorbing dielectric 
insulators~\cite{Barnett}, to the case of correlated metallic systems.
The real, $\epsilon_1(\omega_0,U)=\eta^2(\omega_0,U)-\kappa^2(\omega_0,U)$, and imaginary, 
$\epsilon_2(\omega_0,U)=2\eta(\omega_0,U)\kappa(\omega_0,U)$, parts of the relative permitivity are
given in terms of the refraction index, $\eta(\omega,U)$, and extinction coefficient, $\kappa(\omega,U)$, 
of the correlated metal. The integral over $k$ was done by extending the integration from 
$[0,\infty)$ to $(-\infty,\infty)$, and closing the contour of integration in the upper-half of the complex $k$ 
plane. Note that, since $\eta(\omega_0,U)< 1$ inside the metal, the spontaneous emission rate $\Gamma_{i}$ 
for the $i$-th isolated O-vac is smaller than the result in free space $\Gamma_0$ and the 
role of $\eta(\omega_0,U)\neq 0$, which encodes electronic correlations, is to allow for transmission 
in the visible, forbidden for $\omega_0<\omega_{pl}(0)$, {\it i.e.}, in the absence of correlations.

The highly diffusive character of O-vacs in oxides, which naturally occur even in the purest ABO$_3$ samples,
allows for their migration and easy accumulation near grain boundaries \cite{Schie2012}, their binding 
to foreign dopants \cite{Arora2017}, or their segregation at other defects \cite{Nyman2012}. 
Alternatively, O-vac rich regions can be engineered in a controlled way via microwave irradiation 
\cite{Microwave-Irradiation} or pulsed laser deposition \cite{PLD}, forming thermodynamically 
stable agglomerates of micrometer sizes containing a large number of color centers. In order to model 
such agglomerates we consider a collection of electric-dipoles, $\vec{\mu}$, enclosed by a spherical cavity of 
radius $a$, which acts as a boundary between a region rich in O-vacs and the bulk metal, 
Fig. \ref{Fig-Local-Fields-Cavity}. The local 
fields on a given dipole, due to the presence of all other dipoles, is accounted for by the usual Lorentz local-field 
factor, ${\cal{L}}_{cav}=\left|(\varepsilon(\omega_0)+2)/3\right|$ \cite{Lorentz}. Most importantly, spontaneous 
emission will now occur due to the coupling to the cavity's electromagnetic modes.
The dipole-rich region can be seen 
as an inclusion, characterized by a dielectric constant, $\varepsilon_{in}$, and the enclosing metal as a host, 
with dielectric constant, $\varepsilon_{out}$. The mismatch $\varepsilon_{in}\neq\varepsilon_{out}$, allows for 
the reflection of the emitted radiation back into the center of the cavity, transforming the dipole-rich region into 
a Mie resonator. In this case, the SE rate corresponds to the convolution \cite{Haroche} 
\be
\Gamma_{cav}=\Gamma_0\int_0^\infty d\omega
F_{cav}(\omega)\left|\frac{\varepsilon_{in}(\omega)+2}{3}\right|^2\eta_{in}(\omega)\,\delta(\omega-\omega_0),
\ee
where the Purcell cavity-enhancement factor $F_{cav}(\omega)$ is \cite{Bonod}
\be
F_{cav}(\omega)=\frac{3\pi c^3}{\omega}\sum_{q=1}^\infty{\cal{I}}m\left[\frac{1}{V_{1,0,q}\omega_{1,0,q}(\omega_{1,0,q}-\omega)}\right],
\ee
and $V_{j,m_z,q}$ are the mode volumes for the $(j,m_z,q)$ cavity modes corresponding to the complex 
valued eigenfrequencies $\omega_{j,m_z,q}=\omega^\prime_{j,m_z,q}+i\omega^{\prime\prime}_{j,m_z,q}$.
For dipoles along the $\hat{z}$ direction only the $j=1$ and $m_z=0$ contributions are relevant and 
the associated frequencies are solutions to the equation
\begin{equation}
\sqrt{\varepsilon_{in}}\psi_{1}(k_{in}a)\xi^\prime_{1}(k_{out}a)-\sqrt{\varepsilon_{out}}\psi^\prime_{1}(k_{in}a)\xi_{1}(k_{out}a)=0,
\end{equation}
where $k_{in,out}=\omega/c_{in,out}$, with $c_{in,out}$ being the speed of light inside and 
outside the cavity, and $\psi_1(x)=xj_1(x)$ and $\xi_1=xh_1^{(1)}(x)$, are written in terms
of the $j_1(x)$ spherical Bessel function and the $h_1^{(1)}=j_1(x)+iy_1(x)$ spherical Hankel 
funcion, with the prime indicating derivative with respect to its argument. The mode volumes, 
$V_{1,0,q}$, need not be identical to the physical cavity volume, 
$V_{cav}=4\pi a^3/3$, and, in fact, are a decreasing function of $q$ \cite{Bonod}. 
The cavity modes are, in turn, characterized by a discrete set of frequencies,
$\omega^\prime_{1,0,q}$, that increase with $q=1,2,\dots$, and inverse lifetimes, 
$\omega^{\prime\prime}_{1,0,q}$, that decrease with $q=1,2,\dots$ \cite{Bonod}.

\begin{figure}
\includegraphics[scale=0.34]{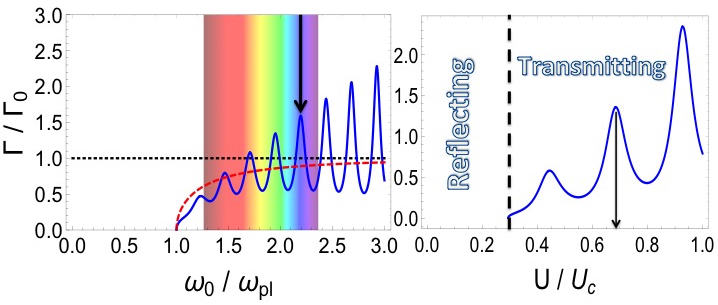}
\caption{Left: spontaneous emission rate, $\Gamma/\Gamma_0$, as a function of the 
emitter's frequency, $\omega_0/\omega_{pl}$. Black arrows at $\omega_0=2.2\omega_{pl}$ 
and $U/U_c=0.67$ correspond to SrVO$_3$. The red (dashed) line is for dilute, isolated emitters in a 
nearly lossless metal, and the full (blue) line is for a resonant spherical cavity, filled up with 
an agglomerate of emitters, again in a nearly lossless metal. Right: the normalized SE rate 
as a function of the correlation strength, $U$, relative to the critical value, $U_c$, for the Mott  
insulator transition. For values of $U$ corresponding to the situation of {\it on-resonance} the 
SE rate, $\Gamma$, rapidly and significantly increases with respect to the free space value, $\Gamma_0$.}
\label{fig4}
\end{figure}

In Fig.~\ref{fig4} the normalized SE rate $\Gamma/\Gamma_0$  is shown, as a function 
of $\omega_0/\omega_{pl}$ (left) and of $U/U_c$ (right) for: i) a single emitter in a nearly lossless transparent metal 
(red dashed line); and ii) an agglomerate of emitters confined to an O-vac rich spherical 
cavity, regarded as an optically active inclusion in an otherwise transparent metallic host 
(solid blue line). The case of SrVO$_3$, with $U\approx 3.2$eV, $\hbar\omega_{pl}< 1.33$eV 
and $\hbar\omega_0\approx 2.8$eV, is represented by black arrows at 
$\omega_0/\omega_{pl}=2.2$ (left) and $U/U_c=0.67$ (right). The real part of the cavity 
frequencies, $\omega^\prime_{1,0,q}$, are determined by the ratios: i) $\varepsilon_{in}/\varepsilon_{out}\neq 1$; 
and ii) $k_{in}a\sim 1$. If the emitter's frequency, $\omega_0$, coincides with one of the
cavity modes' frequencies, $\omega^\prime_{1,0,q}$, {\it i.e.}, if they are {\it on-resonance}, 
then spontaneous emission can be strongly enhanced. On the other hand,
if the emitter and the cavity modes are
{\it off-resonance}, then spontaneous emission is strongly suppressed to values 
even smaller than half the one in free space, see Fig.~\ref{fig4}. If we recall that 
$\omega_0(t,V,U)$ and $\omega_{pl}(U)$ are functions of $t,V$ and $U$, it is 
clear that, even the smallest variations in any of such parameters, especially $U$, could
switch the on- and off- resonance situations, causing the emitter to blink in a controlled way. 

%
%

A number of strategies are known to dynamically vary the values of the parameters $t,V$ and $U$,  
thus allowing for the tuning of the resonances in a controlled way. For instance, the Hubbard $U$ 
parameter can be decreased by almost $10\%$ on femtosecond timescales via laser driving in 
correlated materials~\cite{dejean}. Dynamically decreasing $U$ would not only red-shift the emitter's
frequency $\omega_0(t,V,U)$, but would, at the same time, blue-shift the plasmon frequency 
$\omega_{pl}(U)$, producing, simultaneously, an off-resonance situation while reducing 
the refraction index of the material, $\eta(U)\rightarrow 0$. Alternatively, with $U$ fixed one 
could also use mechanical strain, normal, $\varepsilon_{ij}$, or shear, $\gamma_{ij}$, with $i,j=x,y,z$, 
to control spontaneous emission. In ABO$_3$ oxides strain can be used to modify both the B-O-B 
and B-B bond angles, through the rotation of the oxygen octahedra \cite{Rotation-Octahedra}, 
which leads to changes in the hopping parameter $t$ and positions of the $e_g$ levels. In this 
case, although the plasmon, $\omega_{pl}$, and cavity mode, $\omega_{1,0,q}$, frequencies 
are kept unchanged, the emitter's frequency, $\omega_0(t,V,U)$, can be tuned via strain, which 
modifies $t(\varepsilon_{ij},\gamma_{ij})$. 

In conclusion, we have investigated spontaneous emission in transparent metals subject to electronic correlations.  
We have demonstrated that there exists a critical value for the correlation 
strength $U$, above which spontaneous emission is not only allowed but is also strongly enhanced due to  
resonant coupling between the emitter's
electric dipole moment to long-lived electromagnetic modes inside an optically 
cavity filled up with two-level color centers (oxygen vacancies agglomerates). The situations 
{\it on- and off-} resonance can be tuned in a controlled way either through variations of 
the correlation strength and/or by applying mechanical strain/stress, thus strongly enhancing 
or suppressing spontaneous emission. Altogether, our results suggest concrete and feasible 
routes towards the external control of spontaneous emission in metallic systems that may be applied to solid state single
photon sources to produce photons on demand.

\begin{acknowledgments}
We acknowledge CNPq, CAPES, and FAPERJ for financial support. F.A.P. also thanks the The 
Royal Society-Newton Advanced Fellowship (Grant no. NA150208) for financial support.
\end{acknowledgments}

\end{document}